# Spin-Conserving Resonant Tunneling in Twist-Controlled WSe$_2$-hBN-WSe$_2$ Heterostructures

Kyounghwan Kim,[1] Nitin Prasad,[1] Hema C. P. Movva,[1] G. William Burg,[1] Yimeng Wang,[1] Stefano Larentis,[1] Takashi Taniguchi,[2] Kenji Watanabe,[2] Leonard F. Register,[1] Emanuel Tutuc[1,*]

[1]Microelectronics Research Center, Department of Electrical and Computer Engineering,

The University of Texas at Austin, Austin, TX 78758, USA

[2]National Institute for Materials Science, 1-1 Namiki Tsukuba, Ibaraki 305-0044, Japan

*E-mail: etutuc@mail.utexas.edu, Phone: 512-471-4960

ABSTRACT: We investigate interlayer tunneling in heterostructures consisting of two tungsten diselenide ($WSe_2$) monolayers with controlled rotational alignment, and separated by hexagonal boron nitride. In samples where the two $WSe_2$ monolayers are rotationally aligned we observe resonant tunneling, manifested by a large conductance and negative differential resistance in the vicinity of zero interlayer bias, which stem from energy- and momentum-conserving tunneling. Because the spin–orbit coupling leads to coupled spin-valley degrees of freedom, the twist between the two $WSe_2$ monolayers allows us to probe the conservation of spin-valley degree of freedom in tunneling. In heterostructures where the two $WSe_2$ monolayers have a 180° relative twist, such that the Brillouin zone of one layer is aligned with the time-reversed Brillouin zone of the opposite layer, the resonant tunneling between the layers is suppressed. These findings provide evidence that in addition to momentum, the spin-valley degree of freedom is also conserved in vertical transport.

In recent years, van der Waals (vdW) heterostructures consisting of two-dimensional (2D) materials have been subject to increased research scrutiny[1]. The extensive and continually increasing palette of available 2D materials renders vdW heterostructures attractive for both fundamental science and emerging device applications. Notable examples include moiré patterns in rotationally aligned graphene-on-hexagonal boron nitride (hBN) heterostructures and the emergence of a Hofstadter butterfly in their energy spectrum in high magnetic fields[2–4], moiré patterns in small-twist-angle bilayer graphene[5–7], which exhibit gate-tunable Mott insulators and superconductivity at zero magnetic fields[8,9], and interaction-induced broken symmetry states in high magnetic fields[7].

An attribute essential to semiconductor heterostructures' device functionality, but which remains largely unexplored for most vdW heterostructures, is the coupling and transport along the vertical axis. Interlayer momentum-conserving (resonant) tunneling in rotationally aligned vdW heterostructures may enable novel device functionality for beyond CMOS low-power, high-speed logic[10–13], and resonant tunneling in double layers separated by a tunnel barrier provides a direct measurement of interlayer coupling and the quantum state lifetime[14]. Recent progress in realization of twist-controlled vdW heterostructures[2–4,6,7,15] with precise rotational alignment between 2D layers opens interesting avenues to probe new physics and device functionalities. Negative differential resistance (NDR) characteristics associated with momentum-conserving tunneling have been reported in graphene-based double layers, such as rotationally aligned double monolayer[16,17], and double bilayer graphene[18,19] separated by hBN, or double bilayer graphene separated by a tungsten diselenide ($WSe_2$) tunnel barrier[20]. Theoretical considerations dictate that transition metal dichalcogenide (TMD) double layers can exhibit gate-tunable NDR with very narrow resonances thanks to the increased density of states, and consequently the quantum capacitance[21].

Tungsten diselenide is a prototypical TMD with a honeycomb lattice, which can be isolated down to a monolayer using micromechanical exfoliation, and has high intrinsic mobility at low temperatures[22]. In the monolayer limit, the band extrema are located at the corners (*K* point) of the hexagonal Brillouin zone with a band-gap of approximately 2.0 eV[23,24]. The strong spin−orbit coupling leads to a large valence band splitting of approximately 0.5 eV, with each of the valence bands at a Brillouin zone corner possessing opposite spin[25,26]. In this study, we demonstrate resonant tunneling in dual-gated, rotationally aligned double monolayer $WSe_2$ heterostructures separated by an interlayer hBN dielectric, which reveal narrow tunneling resonances with intrinsic broadening of 1–3 meV at low temperatures. Remarkably, the resonant tunneling is present (absent) in samples where the relative twist between the two WSe$_2$ monolayers is an even (odd) multiple of 60°, a finding which can only be explained by the conservation of the spin-valley degree of freedom in tunneling.

Because monolayer WSe$_2$ band extrema are located at the $K$ point, momentum-conserving tunneling occurs if the relative twist between the two WSe$_2$ monolayers is a multiple of 60°. If the relative twist is an even multiple of 60° (e.g., 0°), the $K$ point and its time-reversed partner ($K'$) of the two monolayers are aligned in momentum space. Conversely, if the relative twist is an odd multiple of 60° (e.g., 180°), the $K$ ($K'$) point of one layer is aligned with the time reversed partner $K'$ ($K$) of the opposite layer. We employ temperature dependent, four-point current–voltage measurements to probe the intrinsic tunneling characteristics at zero and high magnetic fields, and investigate the impact of energy, momentum, and spin conservation.

Figures 1(a) and 1(b) show optical micrographs during the fabrication process of a WSe$_2$-hBN-WSe$_2$ sample (Device #1). First, a large single grain WSe$_2$ monolayer, obtained using either mechanical exfoliation or a recent Au exfoliation technique[27] is trimmed into two smaller flakes using electron-beam lithography and CHF$_3$ plasma etching. The two resulting flakes serve as the top-layer (TL WSe$_2$) and bottom-layer (BL WSe$_2$). In parallel, hBN top-gate (T-hBN) and interlayer (IL-hBN) dielectrics are identified, and the flakes assembled into a (T-hBN)-(TL WSe$_2$)-(IL-hBN)-(BL WSe$_2$) heterostructure, while controlling the relative rotational alignment between the two WSe$_2$ layers[15]. The stack is then placed on prepatterned bottom Pt contacts on an hBN bottom-gate (B-hBN) dielectric on top of a bottom Pt back-gate. Lastly, a Pd top-gate is patterned to cover both the top and bottom WSe$_2$ flakes. Figure 1(c) and (d) show cross sectional schematics of the final device configuration and the biasing scheme, respectively. The Supporting Information [sections S1 and S2] provides details of the top-gate, back-gate, and interlayer capacitance values of the devices discussed in this study, and specifics of the fabrication process, including techniques to accurately control the twist angle between the top and bottom WSe$_2$ flakes. Five devices were investigated in this study, all with consistent results. We focus here on data from three devices, labeled #1–3.

The devices are characterized at negative top-gate ($V_{TG}$) and back-gate ($V_{BG}$) biases in order to populate both the WSe$_2$ layers with holes. The bottom Pt contacts in combination with negative $V_{TG}$ ensure Ohmic *p*-type contacts to WSe$_2$[28]. The intrinsic tunneling characteristics, free of contact resistance effects can be studied, thanks to multiple contacts on each layer. The devices are probed by applying an interlayer voltage ($V_{IL}$) split symmetrically across the two layers, i.e., +$V_{IL}$/2 on the top-layer and –$V_{IL}$/2 on the bottom-layer, and measuring the interlayer bias ($\Delta V_{IL}$) using additional contacts on each layer. Figure 1(e) shows the room temperature two-point (four-point) interlayer tunneling current, $I_{IL}$, vs $V_{IL}$ ($\Delta V_{IL}$), for Device #1 with a three-monolayer thick IL-hBN, where the two WSe$_2$ layers are rotationally aligned with respect to each other, i.e., 0º twist. A salient feature of Figure 1(e) data is the NDR characteristic on either side of $V_{IL}$ = 0 V, or $\Delta V_{IL}$ = 0 V, the telltale sign of resonant tunneling between two 2D carrier systems.

A clear difference between the two-point and four-point data of Figure 1(e) is the apparent stretching of the two-point $I_{IL}$ vs $V_{IL}$ along the $V_{IL}$ axis in comparison to the four-point $I_{IL}$ vs $\Delta V_{IL}$. This difference is due to the contact resistances, which drop a substantial portion of the applied $V_{IL}$, resulting in a reduced $\Delta V_{IL}$. Figure 1(f) shows a similar set of data for Device #2 with a four-monolayer thick IL-hBN, which also shows NDR around $V_{IL}$ = 0 V, albeit with a reduced $I_{IL}$ as expected for a thicker IL-hBN. The difference between the two-point and the four-point data is smaller in Figure 1(e), owing to a larger interlayer tunneling resistance compared to the contact resistance. The data of both Figures 1(e) and 1(f) also show an inflexion in $I_{IL}$ at $\Delta V_{IL} \approx \pm 0.5$ V. We tentatively attribute this feature to the tunneling contributions from the second valence band of WSe$_2$. In the subsequent discussion, we focus only on four-point tunneling data.

Figure 2(a) shows a schematic of the valence bands in the two WSe$_2$ layers at 0° twist. The $K$ and $K'$ valleys in both layers are aligned in momentum space. Figures 2(b)–(d) show the band alignments under an applied $\Delta V_{IL} = 0$ V, $\pm 2\lambda$, where $2\lambda \approx 0.5$ V is the spin–orbit coupling induced band splitting at the $K$ and $K'$ points[25,26]. It can be seen that at $\Delta V_{IL} = 0$ V ($\Delta V_{IL} \approx \pm 2\lambda$) states with same (opposite) spins in the two layers are energetically aligned at both $K$ and $K'$. Figure 2(e) shows $I_{IL}$ vs $\Delta V_{IL}$ and the corresponding differential conductance, $dI_{IL}/d\Delta V_{IL}$ vs $\Delta V_{IL}$ for Device #2 at a temperature $T = 1.5$ K, and at three different biasing conditions chosen such that the hole densities ($p$) in the two WSe$_2$ layers are equal and range from $p = 5.0\times10^{12}$ cm$^{-2}$ to $p = 5.6\times10^{12}$ cm$^{-2}$. The $dI_{IL}/d\Delta V_{IL}$ data were obtained by computing the numerical derivative of the measured $I_{IL}$ vs $\Delta V_{IL}$ data. All three traces show a steep NDR and a corresponding sharp differential conductance peak with a narrow full-width at half maximum of 7 mV close to $\Delta V_{IL} = 0$ V. A close-up view of the $I_{IL}$ vs $\Delta V_{IL}$ data at $p = 5.6\times10^{12}$ cm$^{-2}$ is shown in the inset of Figure 2(e). The symmetric shape of $I_{IL}$ and $dI_{IL}/d\Delta V_{IL}$ vs $\Delta V_{IL}$ data around $\Delta V_{IL} = 0$ V is a signature of equal densities in the two WSe$_2$ layers, because a layer density imbalance would cause the differential conductance peak to shift from $\Delta V_{IL} = 0$ V [Supporting Information, section S3]. The discontinuity in the plot near $\Delta V_{IL} = 0$ V is common for NDR devices, and is explained by a larger external, contact and nonoverlapped WSe$_2$ layer resistance in series with a lower negative differential tunneling resistance, which prevents a measurement of $I_{IL}$ in this regime[19,20]. Furthermore, all three traces also show an $I_{IL}$ inflexion at $\Delta V_{IL} \approx \pm 0.5$ V, albeit of varying intensity, which appears as a corresponding differential conductance peak. In the following, we refer to the differential conductance peak at $\Delta V_{IL} = 0$ V as the primary resonance, and at $\Delta V_{IL} \approx \pm 0.5$ V as the secondary resonances.

To investigate the origin of the tunneling resonances, we perform temperature-dependent measurements. Figure 2(f) and 2(g) show $I_{IL}$ vs $\Delta V_{IL}$ and the corresponding $dI_{IL}/d\Delta V_{IL}$ vs $\Delta V_{IL}$ data as a function of varying temperature for Device #2 at $p = 5.6 \times 10^{12}$ cm$^{-2}$. Two noteworthy observations can be made based on these data. First, the NDR associated with the primary resonance becomes more prominent with decreasing temperature. Equivalently, the primary resonance peak becomes sharper and increases in conductance, along with the neighboring dips which become deeper. Second, the inflexion associated with the secondary resonance, and correspondingly its amplitude, decreases with decreasing temperature. The opposite temperature dependences of the primary resonance and secondary resonance suggest a difference in their mechanism of origin.

To gain insight into the tunneling mechanisms at play in our devices, we start by modeling the interlayer current of the WSe$_2$-hBN-WSe$_2$ system using a perturbative Hamiltonian approach[20,29,30]. The band structures of the top [$\epsilon_T(\mathbf{k}, \sigma)$] and bottom [$\epsilon_B(\mathbf{k}, \sigma)$] WSe$_2$ layers as a function of the crystal momentum $\mathbf{k}$ and spin $\sigma$ are computed using a nearest neighbor tight-binding model[31]. The model includes spin−orbit coupling which gives rise to a pure spin-up band in each layer, and a pure spin-down band with their respective bands edges at the opposing $K$ and $K'$ points[31,32].

The electrostatic potential and band alignment of each WSe$_2$ layer is self-consistently computed using the following set of charge-balance equations,

$$C_{IL}\left(\frac{\phi_B}{e} - \frac{\phi_T}{e}\right) - C_T\left(V_{TG} + \frac{\phi_T}{e}\right) = Q_T(\epsilon_T, \mu_T, \phi_T),$$

$$C_{IL}\left(\frac{\phi_T}{e} - \frac{\phi_B}{e}\right) - C_B\left(V_{BG} + \frac{\phi_B}{e}\right) = Q_B(\epsilon_B, \mu_B, \phi_B) \quad (1)$$

where $C_{IL}$ is the interlayer capacitance per unit area, $C_T$ ($C_B$) is the top-gate (bottom-gate) capacitance per unit area, $\mu_T$ ($\mu_B$) is the chemical potential, $\phi_T$ ($\phi_B$) is the electrostatic potential, $Q_T$ ($Q_B$) is the excess charge density of the top (bottom) WSe$_2$ layer, and $e$ is the electron charge. At zero gate biases and zero interlayer voltage, the chemical potentials of both the top- and bottom-layers are assumed to align with the WSe$_2$ monolayer midgap. The excess charge densities $Q_T$ and $Q_B$ are given by

$$Q_{T/B} = -e \sum_{\mathbf{k},\sigma} \left(f\big(\epsilon_{T/B}(\mathbf{k},\sigma) + \phi_{T/B} - \mu_{T/B}\big) - f(E)\right) \quad (2)$$

where $f(E)$ is the Fermi distribution function.

The single particle tunneling current ($I$) between the two WSe$_2$ layers is modeled as

$$I = -e \int_{-\infty}^{\infty} T(E)\big(f(E - \mu_T) - f(E - \mu_B)\big) dE \quad (3)$$

The vertical transmission rate, $T(E)$, of an electron at energy $E$ is given by,

$$T(E) = \frac{2\pi}{\hbar} \sum_{l\mathbf{k}_T\mathbf{k}_B\sigma_T\sigma_B} \left|t_{l\mathbf{k}_T\mathbf{k}_B\sigma_T\sigma_B}\right|^2 A_{TL}(\mathbf{k}_T, \sigma_T, E) A_{BL}(\mathbf{k}_B, \sigma_B, E) \quad (4)$$

where $l$ labels different contributing processes, and $A_{TL}$ ($A_{BL}$) is the spectral density function of the energy states of the top (bottom) WSe$_2$ layer, and $\hbar$ is the reduced Planck constant. The spectral densities are taken to be Lorentzian in form,

$$A(\mathbf{k}, \sigma, E) = \frac{1}{\pi} \frac{\Gamma}{(E - \epsilon(\mathbf{k}, \sigma))^2 + \Gamma^2} \quad (5)$$

where the parameter $\Gamma$ represents the energy broadening of the quasiparticle states. A coherent momentum- and spin-conserving contribution to the interlayer current is modeled as $t_{\mathbf{k}_T \mathbf{k}_B \sigma_T \sigma_B} \propto \delta_{\mathbf{k}_T \mathbf{k}_B} \delta_{\sigma_T \sigma_B}$. A momentum-conserving, but spin-randomizing tunneling contribution is modeled simply as $t_{\mathbf{k}_T \mathbf{k}_B \sigma_T \sigma_B} \propto \delta_{\mathbf{k}_T \mathbf{k}_B}$. The summation is performed over all momentum states in the Brillouin zone, and over the first two valence bands with opposite spins. For a momentum-randomizing tunneling process, Equations (3)-(5) reduce to

$$I \propto \int_{-\infty}^{\infty} g_T(E) g_B(E) \big(f(E - \mu_T) - f(E - \mu_B)\big) dE \quad (6)$$

where $g_T(E)$ ($g_B(E)$) is the spectrally broadened density of states of the top (bottom) layer at energy $E$. The phenomenological dependence in Equation (6) captures both spin-conserving and spin-flipping tunneling, and does not distinguish between the two cases. The strengths of these processes (*l*) are free parameters, along with the energy broadening parameter ($\Gamma$).

Figure 3(a) shows a comparison of $I_{IL}$ vs $\Delta V_{IL}$ calculated from theory and the experimental data of Figure 2 at $p = 5.6 \times 10^{12}$ cm$^{-2}$ and $T = 1.5$ K, and Figure 3(b) shows a close-up of Figure 3(a) data around $\Delta V_{IL} = 0$ V. To assess the contributions of the different tunneling mechanisms to the total experimentally measured $I_{IL}$, the current is calculated by first assuming energy (*E*), momentum (**k**), and spin ($\sigma$) conservation (labeled *E*,**k**,$\sigma$), then by relaxing the constraints on spin conservation (labeled *E*,**k**,$\sigma̸$), and finally by relaxing momentum conservation (labeled *E*,**k̸**). First, we consider the scenario when tunneling is assumed to occur only when momentum and spin conservation are satisfied (*E*,**k**,$\sigma$). The calculated $I_{IL}$ vs $\Delta V_{IL}$ under these conditions are able to reproduce the experimental primary resonance NDR accurately. However, away from $\Delta V_{IL} = 0$ V, the experimental data diverge from calculations, which predict no tunneling current away from the primary resonance, suggesting there must be another mechanism contributing to the tunneling current.

Given the presence of prominent secondary resonance-like features in our experimental data, it is instructive to relax the spin conservation constraint in our calculations, to consider both spin-conserving and spin-flipping tunneling with equal weight ($E$,**k**,$\sigma$). In this case, the calculated tunneling current reproduces the NDR at $\Delta V_{IL}$ = 0 V just as the spin-conserving model, and additionally shows weaker NDR features at $\Delta V_{IL} \approx \pm 0.5$ V due to secondary resonances. This simulated behavior is similar to observations in double bilayer graphene separated by hBN[15,19], where the secondary resonances appear due to tunneling between the lower and higher spin-degenerate sub-bands of bilayer graphene. We note that the calculated $I_{IL}$ decrease with respect to $\Delta V_{IL}$ beyond the secondary resonances is weaker compared to the primary resonance. Around the secondary resonance, the applied $\Delta V_{IL}$ depletes one of the $WSe_2$ layers, leading to a near zero quantum capacitance in that layer. When one of the layers is depleted, the relative alignment of the two $WSe_2$ layers' bands remains almost constant as the applied bias is increased. This leads to a weak dependence of $I_{IL}$ on $\Delta V_{IL}$, and consequently an apparent stretch-out of the secondary resonance. Because the experimental data in Figures 2 and 3 does not show NDR around $\Delta V_{IL} \approx \pm 0.5$ V, and the $I_{IL}$ value greatly exceeds the calculated current at the secondary resonance, we conclude that spin-relaxing, momentum-conserving tunneling is not a dominant tunneling mechanism causing the apparent secondary resonance peaks.

Next, we consider the case when momentum conservation requirement is relaxed, and the tunneling current is proportional to the product of density of states in both the $WSe_2$ layers (labeled $E$,**k**). This model captures the case when the spin is either conserved, or randomized in tunneling.

In this scenario, the calculated current does not show the primary resonance, but does reproduce the experimental data at high $|\Delta V_{IL}|$ reasonably well. Therefore, the interlayer current at high $|\Delta V_{IL}|$ can be attributed predominantly to the momentum-relaxing tunneling mechanisms. Furthermore, the current saturates with a weak increase after a particular $\Delta V_{IL}$, when one of the WSe$_2$ layers is depleted.

Finally, we consider a scenario where both momentum- and spin-conserving, as well as momentum-randomizing tunneling processes are present (labeled $E,\mathbf{k},\sigma + E,\mathbf{k}$). The calculated tunneling current in this case reproduces both the primary resonance and the secondary resonance features of our experimental data, suggesting that both of these mechanisms are simultaneously present in our samples.

We now discuss the temperature dependence of individual layer quasiparticle state broadening determined from the tunneling characteristics. Figure 3(c) shows the value of Γ used to fit the experimental data of both Devices #1 and #2 as a function of temperature. For comparison, Figure 3(c) also shows the broadening associated with the transport scattering time, $\hbar/(2\tau_{tr})$, where $\tau_{tr} = \mu_{tr} m^*/e$ is the transport scattering time, and $\mu_{tr}$ is the hole mobility measured in a separate hBN-encapsulated monolayer WSe$_2$ device. We use $m^* = 0.45 m_e$ as the effective mass of WSe$_2$ holes in the $K$, $K'$ valleys[22], where $m_e$ is the free electron mass. The $\hbar/(2\tau_{tr})$ value shows a noticeable increase with temperature for $T > 50$ K, consistent with phonon scattering, but is weakly dependent on temperature for $T < 50$ K, a regime dominated by fixed impurity scattering[28]. The Γ value, on the other hand, increases with temperature at a much higher rate, indicating that scattering has a greater impact on quasiparticle lifetime than what the transport time may otherwise suggest. This difference is expected because of processes such as electron−electron scattering or small-angle electron−phonon scattering, which do not affect the mobility and transport time as the total momentum is conserved or largely unchanged, but can reduce considerably the quasiparticle lifetime leading to an increase in the $\Gamma$ value[14,33]. At the lowest temperatures, however, disorder is the limiting factor which determines the broadening.

To probe the hypothesis that the primary resonance at $\Delta V_{IL} = 0$ V is indeed due to momentum conserving tunneling, we perform measurements in the presence of an in-plane magnetic field ($B_\parallel$), which is, thus, perpendicular to the direction of tunneling. Figure 4(a) shows $dI_{IL}/d\Delta V_{IL}$ vs $\Delta V_{IL}$ near the primary resonance for Device #2 at various magnetic field values, which shows the resonance peak conductance decreasing and the peak width increasing with increasing magnetic field. The effect of $B_\parallel$ is to produce a momentum shift of the tunneling carriers due to the Lorentz force, which thereby causes a momentum mismatch of $(eB_\parallel d)/\hbar$, and consequently a suppression of the resonance peak; $d$ is the separation between the two WSe$_2$ monolayers. Figure 4(b) shows $dI_{IL}/d\Delta V_{IL}$ vs $B_\parallel$ for both Devices #1 and #2, where $dI_{IL}/d\Delta V_{IL}$ decreases with $B_\parallel$. This observation confirms that momentum-conserving tunneling is responsible for the primary resonance at $\Delta V_{IL} = 0$ V.

To further test the role of spin conservation in tunneling, we now turn our focus to interlayer tunneling in the case of a 180° twist between the two WSe$_2$ layers. Figure 5(a) shows a schematic of the valence bands in the two WSe$_2$ layers at 180° twist and Figures 5(b)–(d) show the band alignments under application of $\Delta V_{IL} = 0, \pm 2\lambda$. Here, in contrast to Figures 2(a)–(d) the $K$ valley in one layer is aligned with the $K'$ valley in the opposite layer in momentum space. At $\Delta V_{IL} = 0$ V ($\Delta V_{IL} \approx \pm 2\lambda$) states with opposite (same) spins in the two layers are energetically aligned. Figure 5(e) shows $I_{IL}$ vs $\Delta V_{IL}$ and the corresponding $dI_{IL}/d\Delta V_{IL}$ vs $\Delta V_{IL}$ at varying temperatures for Device #3 with a 180° twist between the WSe$_2$ layers, and a seven-monolayer thick IL-hBN. The top and bottom layer densities are $p = 5.7\times10^{12}$ cm$^{-2}$ and $p = 5.0\times10^{12}$ cm$^{-2}$, respectively. The small difference in layer densities would slightly shift the position of a tunneling resonance as a function of $\Delta V_{IL}$ [Supporting Information, section S3], but would not otherwise affect the resonances. The data show negligible conductance near $\Delta V_{IL} = 0$ V, namely, no primary resonance, but a marked increase in $I_{IL}$ for $|\Delta V_{IL}| > 0.4$ V, i.e., prominent secondary resonances, including an NDR at $\Delta V_{IL} \approx -0.5$ V. The tunneling current magnitude at the secondary resonances is lower compared to Devices #1 and #2 due to the thicker IL-hBN.

Figure 5(f) shows a comparison of the experimental and calculated tunneling currents under the same set of assumptions employed for the 0º twist devices. In the case of energy-, momentum-, and spin-conserving tunneling ($E$,**k**,$\sigma$) the calculated current shows prominent NDR characteristics at $\Delta V_{IL} \approx \pm 0.5$ V, and zero current near $\Delta V_{IL} = 0$ V. The tunneling current near $\Delta V_{IL} = 0$ V is strongly suppressed due to spin mismatch between the opposite layer bands, despite alignment in the momentum space. In the case when only energy and momentum conservation are enforced ($E$,**k**,$\cancel{\sigma}$), the calculated current shows a primary resonance at $\Delta V_{IL} = 0$ V, at variance with experimental data, strongly suggesting that momentum-conserving tunneling is also spin-conserving. The calculated NDR features at $\Delta V_{IL} \approx \pm 0.5$ V are much more prominent than in the experimental data, which is only observed at $\Delta V_{IL} \approx -0.5$ V. Finally, in the case when momentum-randomizing tunneling is considered ($E$,$\cancel{\mathbf{k}}$), the calculation is able to reproduce the experimental features at high $\Delta V_{IL}$. Comparing the calculations with experimental data, the absence of a primary resonance implies both spin and momentum are conserved near $\Delta V_{IL} = 0$ V, while at high interlayer bias, momentum-randomizing tunneling starts to contribute to tunneling, likely associated with a phonon-mediated process.

In conclusion, we demonstrate rotationally controlled WSe$_2$-based heterostructures. We probe experimentally, and explain theoretically, the role of spin-valley conservation in addition to energy and momentum conservation in interlayer tunneling using twist-controlled WSe$_2$-hBN-WSe$_2$ heterostructures. Devices with 0º twist between the two WSe$_2$ monolayers show a sharp resonance near zero interlayer bias, stemming from momentum- and spin-conserving tunneling, and apparent secondary resonances near ±0.5 V stemming from momentum-randomizing tunneling between the spin-split valence bands in the two WSe$_2$ layers. Devices with 180º twist do not show a primary resonance, but do show prominent secondary resonances thanks to alignment of the lower and upper valence bands in the two WSe$_2$ layers at an interlayer bias of ≈ ±0.5 V.

ASSOCIATED CONTENT

**Supporting Information**

S1: Device Information, S2: Fabrication Process, S3: Effect of Density Imbalance.

AUTHOR INFORMATION

**Corresponding Author**

*Email: etutuc@mail.utexas.edu, Phone: 512-471-4960

**Author Contributions**

K.K. fabricated the devices, and performed the experimental measurements, with assistance from H.C.P.M, G.W.B., Y.M., and S.L. K.K., N.P., H.C.P.M., L.R.F., and E.T. analyzed the results. N.P. and L.F.R. performed the theoretical modeling. T.T. and K.W. supplied the hBN crystals. K.K., N.P., H.C.P.M., L.F.R. and E.T. contributed to the manuscript.

**Notes**

The authors declare no competing financial interest.

ACKNOWLEDGEMENTS

This work was supported by the Semiconductor Research Corp. Nanoelectronics Research Initiative SWAN center, National Science Foundation grant EECS-1610008, and Samsung Corp. K.W. and T. T. acknowledge support from the Elemental Strategy Initiative conducted by the MEXT, Japan and JSPS KAKENHI Grant No. JP15K21722.

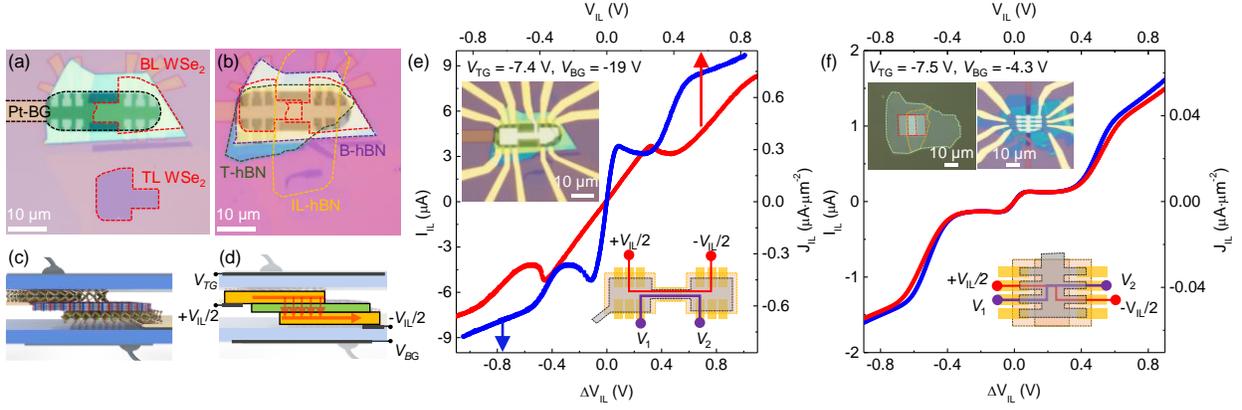

**Figure 1.** Optical micrographs of WSe$_2$-hBN-WSe$_2$ Device #1 (a) during the fabrication process, and (b) after assembly, but before top-gate patterning. The individual layers and the Pt back-gate are outlined by dashed lines. (c) Cross sectional schematic of the final device, and (d) biasing scheme used for the two-point interlayer tunneling current measurement. Two-point $I_{IL}$ vs $V_{IL}$ (top axis), and four-point $I_{IL}$ vs $\Delta V_{IL}$ (bottom axis) for (e) Device #1 and (f) Device #2 at room temperature. The right axes in panels (e) and (f) show $I_{IL}$ normalized to the TL and BL WSe$_2$ overlap area. Top left insets of panels (e) and (f): optical micrographs of (e) Device #1 after fabrication and (f) Device #2 during and after fabrication. Bottom right insets of panels (e) and (f): biasing scheme used for the four-point measurements of (e) Device #1 and (f) Device #2. The four-point interlayer bias is $\Delta V_{IL} = V_1 - V_2$.

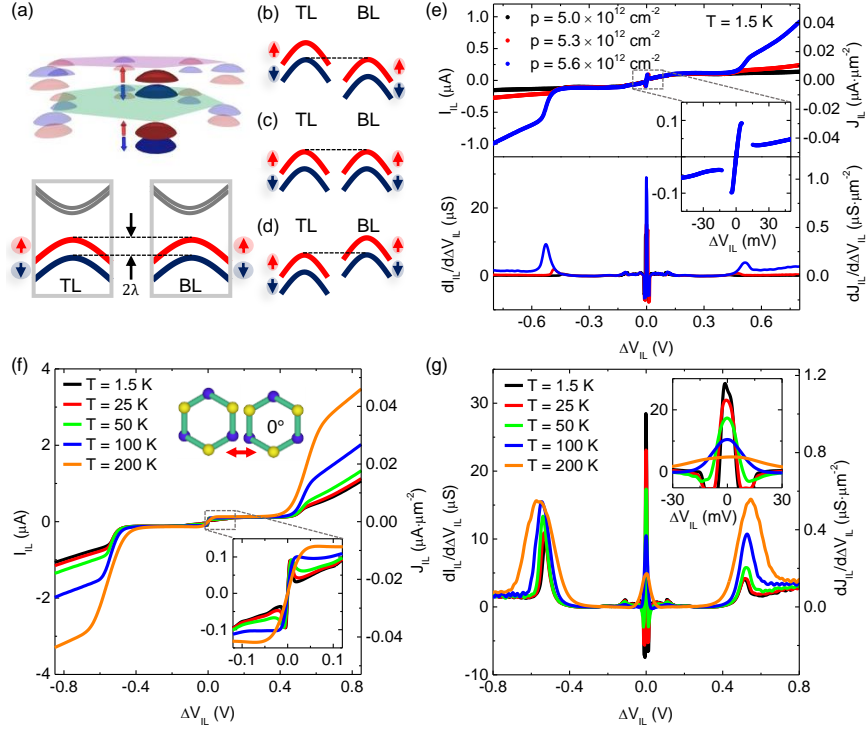

**Figure 2.** (a) Band alignment schematic near the *K* and *K′* points of the two WSe$_2$ layers with 0º twist. Red and blue mark the two spin orientations. (b–d) Band alignments under an applied $\Delta V_{IL}$ = -2λ, 0, 2λ. (e) $I_{IL}$ vs $\Delta V_{IL}$ (top panel) and $dI_{IL}/d\Delta V_{IL}$ vs $\Delta V_{IL}$ (bottom panel) for Device #2 at *T* = 1.5 K measured at equal layer density values, from $p = 5.0 \times 10^{12}$ cm$^{-2}$ to $p = 5.6 \times 10^{12}$ cm$^{-2}$. Inset: close-up of $I_{IL}$ vs $\Delta V_{IL}$ at $p = 5.6 \times 10^{12}$ cm$^{-2}$ near $\Delta V_{IL}$ = 0 V. (f) $I_{IL}$ vs $\Delta V_{IL}$ and (g) $dI_{IL}/d\Delta V_{IL}$ vs $\Delta V_{IL}$ for Device #2 as a function of varying temperature from *T* = 1.5 to 200 K, and $p = 5.6 \times 10^{12}$ cm$^{-2}$. Panel (f) bottom inset and panel (g) top inset are close-up views of the corresponding panel data near $\Delta V_{IL}$ = 0 V. Top inset of panel (f) is an illustration of the 0º twist between the two WSe$_2$ lattices. The right axes of panels (e–g) show $I_{IL}$ and $dI_{IL}/d\Delta V_{IL}$ normalized to the TL and BL WSe$_2$ overlap area.

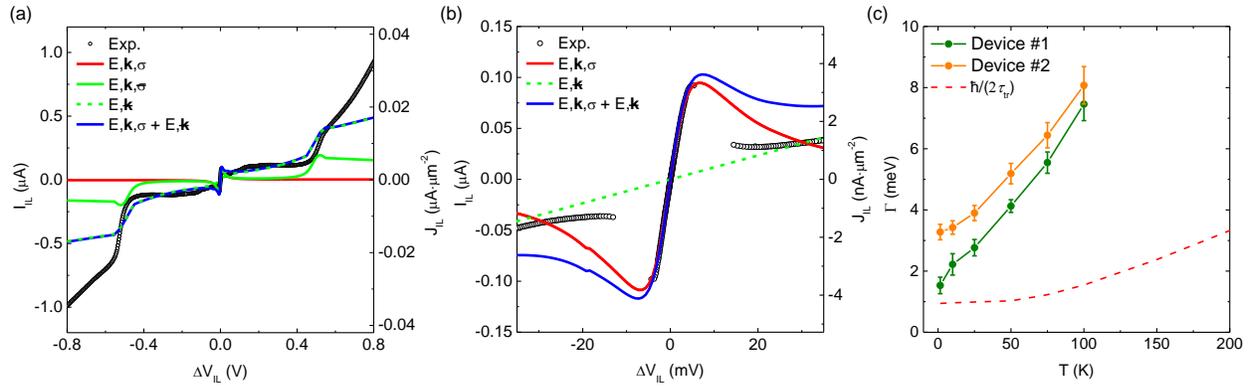

**Figure 3.** (a) Experimental $I_{IL}$ vs $\Delta V_{IL}$ of Device #2 at $p = 5.6 \times 10^{12}$ cm$^{-2}$ and $T = 1.5$ K (circles), and calculations under different combinations of energy-conserving ($E$), momentum-conserving (**k**) or momentum-randomizing (**k̶**), and spin-conserving ($\sigma$) or spin-randomizing ($\sigma\!\!\!/$) tunneling processes contributing to $I_{IL}$. (b) Close-up view of panel (a) near $\Delta V_{IL} = 0$ V. (c) $\Gamma$ vs $T$ for Device #1 and Device #2 extracted from fits to the experimental $I_{IL}$ vs $\Delta V_{IL}$ data (symbols), and from monolayer WSe$_2$ mobility (dashed line) corresponding to the transport lifetime. The right axes of panels (a) and (b) show $I_{IL}$ normalized to the TL and BL WSe$_2$ overlap area.

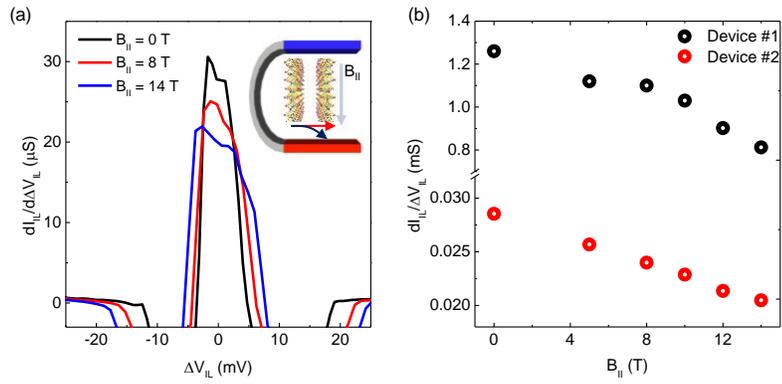

**Figure 4.** (a) $dI_{IL}/d\Delta V_{IL}$ vs $\Delta V_{IL}$ for Device #2 at $T = 1.5$ K near $\Delta V_{IL} = 0$ V measured at $B_\parallel = 0$, 8, and 14 T. Inset: schematic of the sample orientation with respect to $B_\parallel$. (b) Peak $dI_{IL}/d\Delta V_{IL}$ vs $B_\parallel$ for Device #1 and Device #2.

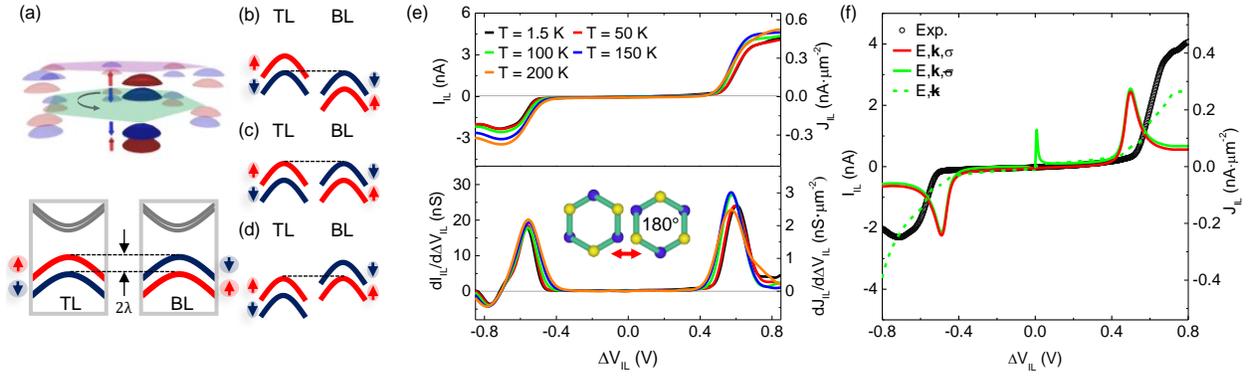

**Figure 5.** (a) Schematic of the band alignment near the $K$ and $K'$ points in the two WSe$_2$ layers at 180º twist. Red and blue mark the two spin orientations. (b–d) Band alignments under an applied $\Delta V_{IL}$ = -2$\lambda$, 0, 2$\lambda$. (e) $I_{IL}$ vs $\Delta V_{IL}$ (top panel) and d$I_{IL}$/d$\Delta V_{IL}$ vs $\Delta V_{IL}$ (bottom panel) for Device #3 as a function of varying temperature from $T$ = 1.5 to 200 K. The top and bottom layer densities are 5.7×10$^{12}$ cm$^{-2}$ and 5.0×10$^{12}$ cm$^{-2}$, respectively. Inset: illustration of the two monolayer WSe$_2$ lattices with 180º twist. (f) $I_{IL}$ vs $\Delta V_{IL}$ of panel (e) data at $T$ = 1.5 K (circles), and calculations under different combinations of tunneling mechanisms. The right axes of panels (e) and (f) show $I_{IL}$ and d$I_{IL}$/d$\Delta V_{IL}$ normalized to the TL and BL WSe$_2$ overlap area.

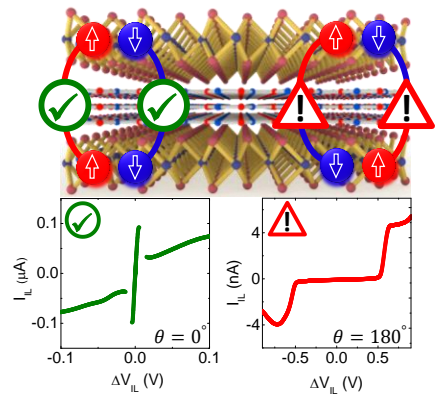

Figure TOC

# Spin-Conserving Resonant Tunneling in Twist-Controlled $WSe_2$-hBN-$WSe_2$ Heterostructures

## Supporting Information

*Kyounghwan Kim,[1] Nitin Prasad,[1] Hema C. P. Movva,[1] G. William Burg,[1] Yimeng Wang,[1] Stefano Larentis,[1] Takashi Taniguchi,[2] Kenji Watanabe,[2] Leonard F. Register,[1] Emanuel Tutuc[1],\**

[1]Microelectronics Research Center, Department of Electrical and Computer Engineering,

The University of Texas at Austin, Austin, TX 78758, USA

[2]National Institute for Materials Science, 1-1 Namiki Tsukuba, Ibaraki 305-0044, Japan

*E-mail: etutuc@mail.utexas.edu, Phone: 512-471-4960

**Contents:**

S1: Device Information

S2: Fabrication Process

S3: Effect of Density Imbalance

## S1: Device Information

Table S1. Summary of the twist angle, T-hBN, B-hBN, and IL-hBN thicknesses, and the top-gate, back-gate, and interlayer capacitances for the three devices investigated in this study.

| Sample | WSe$_2$-WSe$_2$ Twist Angle | T-hBN Thickness | B-hBN Thickness | IL-hBN # of layers | $C_T$ (nF/cm$^2$) | $C_B$ (nF/cm$^2$) | $C_{IL}$ (µF/cm$^2$) |
|---|---|---|---|---|---|---|---|
| Device #1 | 0° | 23.3 nm | 60.3 nm | 3 | 114 | 44 | 2.30 |
| Device #2 | 0° | 23.0 nm | 13.0 nm | 4 | 115 | 204 | 1.80 |
| Device #3 | 180° | 23.8 nm | 29.7 nm | 7 | 112 | 89 | 1.15 |

Table S1 shows the twist angle, top-hBN, bottom-hBN, and IL-hBN thicknesses, and the top-gate ($C_T$), back-gate ($C_B$), and interlayer ($C_{IL}$) capacitance values for the three devices investigated in this study. $C_{TG}$ and $C_{BG}$ are calculated as $C_{T,B} = \epsilon_0 k_{hBN}/t_{hBN}$, where $\epsilon_0$, $k_{hBN}$, and $t_{hBN}$ are the permittivity of free space, hBN dielectric constant, and hBN thickness measured by atomic force microscopy, respectively. We use $k_{hBN} = 3$, a value determined experimentally by from a scaling of measured gate capacitance values vs $t_{hBN}$, in multiple gated WSe$_2$ Hall bars with hBN dielectrics. To better approximate the $C_{IL}$ value at reduced hBN thicknesses, we use $C_{IL}^{-1} = t_{hBN}/\epsilon_0 k_{hBN} + C_0^{-1}$, where $C_0$ includes contributions from the two interfaces. The experimentally measured $C_{IL}$ values in rotationally aligned double bilayer graphene separated by hBN are 3.1 µF/cm$^2$, 1.8 µF/cm$^2$, and 1.5 µF/cm$^2$ for $t_{hBN}$ corresponding to 2, 4, and 5 hBN monolayers[1]. Using these values we determine $C_0 = 9$ µF/cm$^2$.

## S2: Fabrication Process

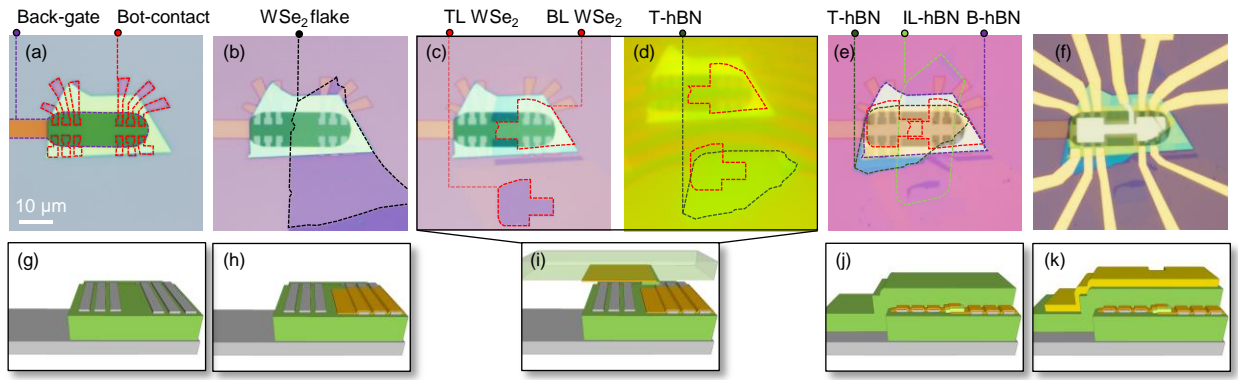

Figure S2. Optical micrographs (a-f) and schematics (g-k) of the detailed fabrication process of a typical WSe$_2$-hBN-WSe$_2$ heterostructure sample (Device #1). Panels (a-f) are at the same magnification.

Figure S2 describes the fabrication process flow of Device #1, a dual-gated WSe$_2$-hBN-WSe$_2$ heterostructure comprising two rotationally aligned monolayer WSe$_2$ electrodes separated by an interlayer hBN tunnel barrier. Figures S2(a-f) show the optical micrographs and Figures S2(g-k) the corresponding schematics during the fabrication process. Using a layer transfer method,[1,2] we first transfer a bottom-gate hBN (B-hBN) flake, which serves as the bottom-gate dielectric, onto a pre-defined metal local back-gate patterned using electron-beam lithography (EBL), electron-beam metal evaporation (EBME) of Cr/Pt (2 nm/8 nm), and lift-off. After the B-hBN is transferred, bottom electrodes of Cr/Pt (3 nm/12 nm) are patterned using EBL and EBME [Figures S2(a, g)], which serve as *p*-type contacts to the WSe$_2$ layers[3]. On a separate SiO$_2$/Si substrate, monolayer WSe$_2$ flakes are exfoliated and confirmed using Raman and photoluminescence (PL) spectroscopy. A suitable monolayer WSe$_2$ flake is chosen and transferred onto the Pt contacts meant for the bottom WSe$_2$ electrode in such a way that only a partial region of the flake is on the contacts, as shown in Figures S2(b, h). We note here that a relatively large area monolayer WSe$_2$ flake is chosen to facilitate trimming in order to obtain both the bottom (BL) and top (TL) layer WSe$_2$ layers from a single crystal grain.

The monolayer WSe$_2$ flake is then sectioned into the individual TL and BL WSe$_2$ electrodes using EBL followed by etching in a CHF$_3$ plasma [Figure S2(c)]. Subsequently, we first pick-up a top-gate hBN flake (T-hBN) from a separate SiO$_2$/Si substrate using a modified half hemispherical polydimethylsiloxane (PDMS) stamp coated with an adhesive polypropylene carbonate (PPC) layer, and then selectively pick-up only the TL WSe$_2$ [Figure S2(d)], followed by a thin interlayer hBN flake (IL-hBN), and place the resulting (T-hBN)-(TL WSe$_2$)-(IL-hBN) stack on the (BL WSe$_2$)-(B-hBN) substrate after precisely controlling the rotational alignment between the TL and BL WSe$_2$ , which in the case of Device #1 is 0º [Figure S2(e, j)].

To improve the accuracy of alignment, we use an *x-y* grid of pre-patterned metal alignment markers on the substrate, spaced apart by 200 μm, as a reference guide. The alignment markers leave an impression on the PPC/PDMS stamp during TL WSe$_2$ pick-up which is used as a reference for the subsequent alignment with the BL WSe$_2$ after IL hBN pick-up. To complete the device fabrication, we perform two final EBL, EBME, and lift-off steps to pattern the Pd (30 nm) top-gate electrode and Cr/Pd/Au (3 nm/25 nm/40 nm) metal contacts and bond pads [Figure S2(f, k)]. To remove polymer residues introduced during the transfer process, we perform an ultra-high vacuum (1×10$^{-9}$ Torr) anneal at 350°C for 2 hours after each transfer step.

S4

## S3: Effect of Density Imbalance

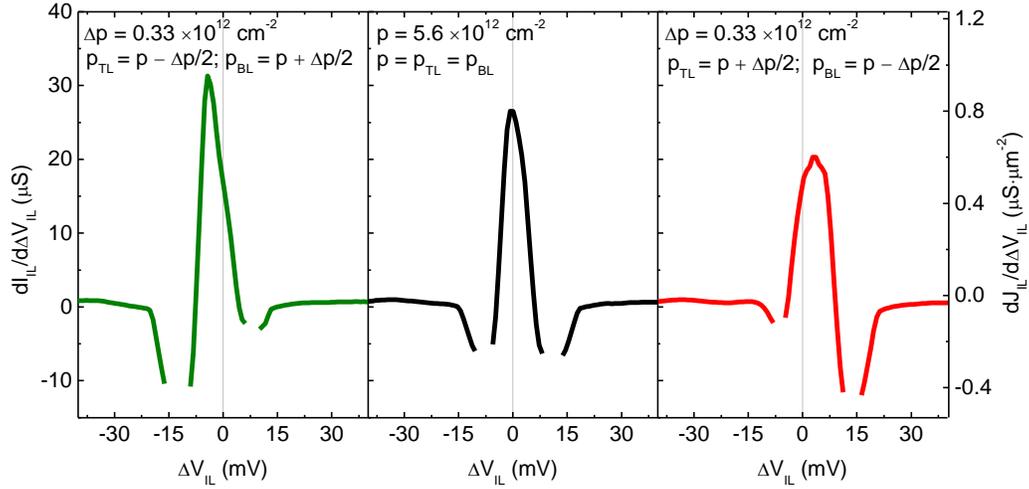

Figure S3. $dI_{IL}/d\Delta V_{IL}$ vs $\Delta V_{IL}$ for Device #2 in the case of balanced (center panel) and imbalanced (left and right panels) TL and BL WSe$_2$ densities. The right axis shows $dI_{IL}/d\Delta V_{IL}$ normalized to the TL and BL WSe$_2$ overlap area.

Figure S3 shows $dI_{IL}/d\Delta V_{IL}$ vs $\Delta V_{IL}$ for Device #2 for three cases where the densities in the TL WSe$_2$ ($p_{TL}$) and BL WSe$_2$ ($p_{BL}$) layers are balanced at $p = p_{TL} = p_{BL} = 5.6\times10^{12}$ cm$^{-2}$ (center panel) and slightly imbalanced, with a density imbalance of $\Delta p = \mp 0.33\times10^{12}$ cm$^{-2}$ (left and right panels). In the case when the two layers have equal densities of holes, the alignment of the bands occurs simultaneously with the alignment of the Fermi levels at $\Delta V_{IL} = 0$ V, leading to a differential conductance peak centered at $\Delta V_{IL} = 0$ V. In the presence of a density imbalance between the two layers, a finite $\Delta V_{IL}$ is needed to align the bands, resulting in a differential conductance peak centered away from $\Delta V_{IL} = 0$ V, depending on the sign and magnitude of the density imbalance[4].